\documentclass[a4paper]{article}
\pdfoutput=1 % if your are submitting a pdflatex (i.e. if you have
\usepackage{upgreek}
\usepackage{graphicx}
\usepackage{jinstpub} % for details on the use of the package, please
\usepackage{hyperref}

\title{The $\upmu$-RWELL layouts for high particle rate}

\author[a,1]{G. Bencivenni,\note{Corresponding author.}}
\author[b]{R. de Oliveira,}
\author[a]{G. Felici,} 
\author[a]{M. Gatta,}
\author[a]{M. Giovannetti,}
\author[a]{G. Morello,}
\author[c]{A. Ochi,}
\author[a]{M. Poli Lener,}
\author[a,d]{E. Tskhadadze} 
\affiliation[a]{Laboratori Nazionali di Frascati dell`INFN,\\Frascati, Italy}
\affiliation[b]{CERN, \\Meyrin, Switzerland}
\affiliation[c]{Kobe University, \\Kobe, Japan}
\affiliation[d]{Technological Institute of Georgia, \\Tbilisi, Georgia}
\emailAdd{giovanni.bencivenni@lnf.infn.it}
\abstract{The $\upmu$-RWELL is a single-amplification stage resistive Micro-Pattern Gaseous Detector (MPGD). The detector amplification element is realized with a single copper-clad polyimide foil micro-patterned with a blind hole (well) matrix and embedded in the readout PCB through a thin Diamond-Like-Carbon (DLC) sputtered resistive film. The introduction of the resistive layer, suppressing the transition from streamer to spark, allows to achieve large gains ($\geq$10$^4$) with a single amplification stage, while partially reducing the capability to stand high particle fluxes.
The simplest resistive layout, designed for low-rate applications, is based on a single-resistive layer with edge grounding. At high particle fluxes this layout suffers of a non-uniform response.
In order to get rid of such a limitation different current evacuation geometries have been designed.
In this work we report the study of the performance of several high rate resistive layouts tested at the CERN H8-SpS and PSI $\uppi$M1 beam test facilities.
These layouts fulfill the requirements for the detectors at the HL-LHC and for the experiments at the next generation colliders FCC-ee/hh and CepC.}

\begin{document}
\maketitle
\flushbottom
\section{Introduction}
\label{sec:intro}

\begin{figure}[h]
	\begin{center}
      	\includegraphics[scale=0.3]{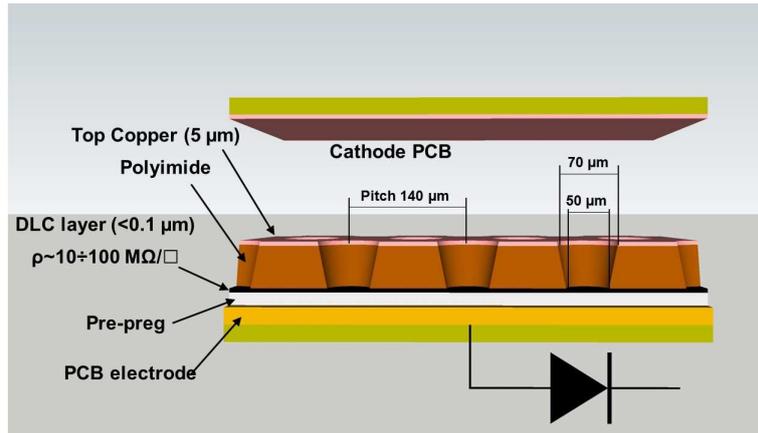}
        \caption{Layout of the $\upmu$-RWELL.}
        \label{rwell-substrate}
    \end{center}
\end{figure}

The  $\upmu$-RWELL, fig.\ref{rwell-substrate}, is a single-amplification stage resistive MPGD \cite{micro-RWELL} that combines in a unique approach the solutions and improvements achieved in the last years in the MPGD field.
The R\&D on $\upmu$-RWELL aims to improve the stability under heavy irradiation while simplifying the construction procedures in view of an easy technology transfer to industry: a milestone for large scale applications in fundamental research at the future colliders and even beyond the HEP.

\noindent The detector is composed of two elements: the cathode, a simple FR4 PCB with a thin copper layer on one side and the $\upmu$-RWELL\_PCB, the core of the detector. The $\upmu$-RWELL\_PCB, a multi-layer circuit realized by means of standard photo-lithography technology, is composed of a well patterned single copper-clad polyimide (Apical{\textsuperscript\textregistered}) foil\footnote{50 $\upmu$m thick polyimide covered on one side with 5 $\upmu$m thick copper, similar to the GEM base material.} acting as amplification element of the detector; a resistive layer, realized with a DLC film sputtered on the bottom side of the polyimide foil, as discharge limitation stage; a standard PCB for readout purposes, segmented as strip, pixel or pad electrodes.

\begin{figure}[h]
	\begin{center}
      	\includegraphics[scale=0.4]{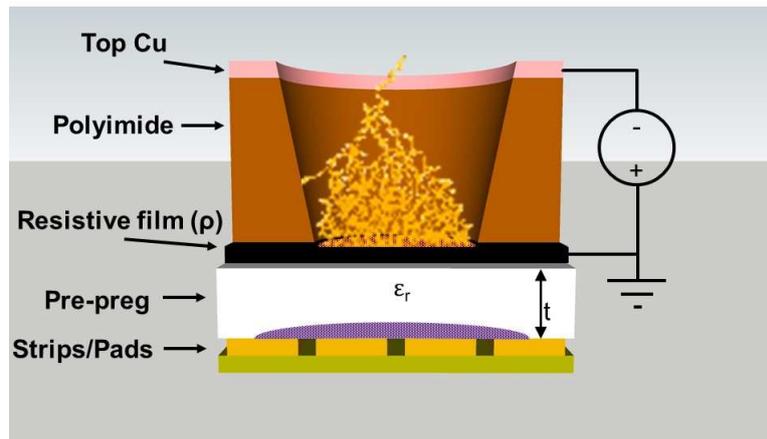}
        \caption{Principle of operation of the $\upmu$-RWELL.}
        \label{principle-operation}
    \end{center}
\end{figure}

\noindent Applying a suitable voltage between the copper layer and the DLC, the well acts as a multiplication channel for the ionization produced in the drift gas gap,  fig.\ref{principle-operation}.
The charge induced on the resistive film is spread with a time constant  \cite{dixit, NIM-RWELL-1}
\begin{displaymath}
\tau=\rho c =\rho\frac{\epsilon_{0}\epsilon_{r}}{t}
\end{displaymath}
being $\rho$ the surface resistivity (in the following simply called resistivity), $c$ the capacitance per unit area and $t$ the distance between the resistive layer and the readout plane.

\noindent The spark suppression mechanism is similar to the one of the Resistive Plate Counters - RPCs, \cite{Pestov, santo, bencivenni, fonte}: the streamer created inside the amplification volume, inducing a large current flowing through the resistive layer, generates a localized drop of the amplifying voltage with an effective quenching of the multiplication process in the gas. This mechanism strongly suppressing the discharge amplitude allows an operation of the detector at large gains ($\geq$10$^4$) with a single amplification stage.

\noindent A drawback, correlated with the Ohmic behaviour of the detector, is the reduced capability to stand high particle fluxes. 
Indeed a detector relying on a simple single-resistive layout suffers at high particle fluxes of a non-uniform response over its surface, more evident as the size of the detector increases. This effect, correlated to the average resistance ($\Omega$) faced by the charge produced in the avalanche, actually depends on the distance between the particle incidence position and the detector grounding line.

\noindent In this paper we discuss the resistive layouts for high rate purposes and their performance in terms of efficiency and rate capability as measured in different beam conditions at the CERN and PSI beam lines.

\section{The Diamond-Like-Carbon}
The DLC sputtering technology is typically used in applications (mechanics, automotive and medical industry) that require surface hardening and reduced abrasive wear.
The sputtered DLC is a class of carbon material that contains both the diamond as well as the graphite structure.

\begin{figure}[h]
	\begin{center}
    	\includegraphics[scale=0.5]{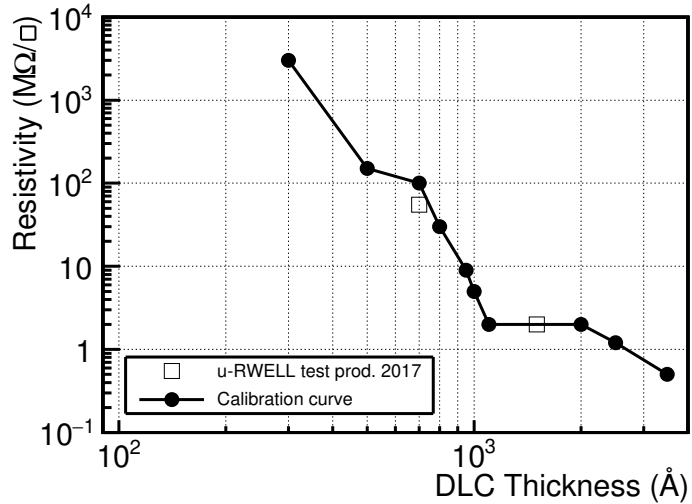}
        \caption{Resistivity as a function of the DLC thickness for different sputtering batches \cite{DLC-Bormio}.}
        \label{DLC}
    \end{center}
\end{figure}

\noindent For the production of our detectors, a large industrial sputtering chamber in Be-Sputter Co., Ltd. (Japan) has been used \cite{DLC}. Starting from a target of graphite, magnetron-sputtered on one side of the Apical{\textsuperscript\textregistered} foil, the DLC film is eventually obtained with suitable surface resistivity and uniform thickness. The resistivity dependence as a function of the DLC thickness for different sputtering batches is shown in fig.~\ref{DLC}. The black calibration curve has been obtained by sputtering the DLC on a brand-new Apical{\textsuperscript\textregistered}  foil, while the open square markers refer to the $\upmu$-RWELL 2017 test production done on a  pre-dried base material (200$^\circ$C for $\sim$2 hours).

\noindent Very recent developments \cite{ZHOU} at USTC-Hefei (PRC) led to the manufacturing of DLC+Cu sputtered Apical{\textsuperscript\textregistered} foils, where additional few microns of copper cover the free surface of the DLC layer. This technology opens the way towards improved high rate $\upmu$-RWELL layouts.

\section{The high rate layouts}
\begin{figure}
	\begin{center}
		\includegraphics[scale=0.3]{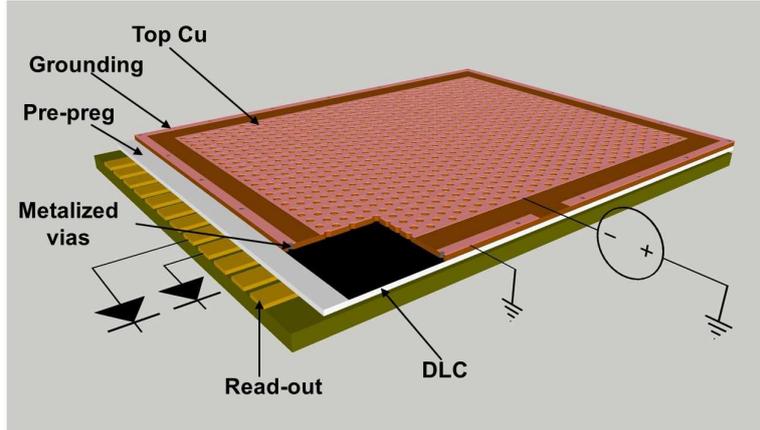}
        \caption{Sketch of the Single-Resistive layout.}
        \label{resistive-stage1}
        \end{center}
\end{figure}

The simplest scheme for the evacuation of the current in a $\upmu$-RWELL is based on a single resistive layer with a grounding line all around the active area, fig.\ref{resistive-stage1}, (Single-Resisitive layout - SRL).
For large area devices the path of the current to ground could therefore be large and strongly dependent on the incidence point of the particle.

\noindent In order to cope with this effect the solution is to reduce as much as possible the average path towards the ground connection, introducing a high density grounding network on the resistive stage.

\noindent Two different high rate (HR) layouts with dense grounding scheme have been implemented: the Double-Resistive Layer (DRL) and the Silver-Grid (SG).
\subsection{The Double-Resistive layout}
\label{sec_DRL}
\begin{figure}
	\begin{center}
		\includegraphics[scale=0.3]{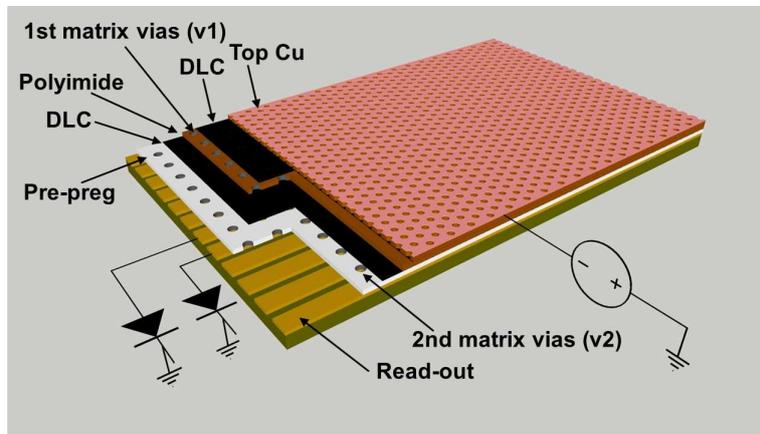}
       \caption{Sketch of the Double-Resistive layout.}
        \label{resistive-stage2}
        \end{center}
\end{figure}

The DRL layout is sketched in fig.~\ref{resistive-stage2}.
The first DLC film, sputtered on the back-plane of the amplification stage, is connected to a second DLC layer by means of a matrix of metalized vias (v1). A further matrix of vias (v2) connects the second DLC film to the underlying readout electrodes, providing the final grounding of the whole resistive stage. The vias density is typically $\leq$ 1 cm$^{-2}$. In this way a 3D-current evacuation layout is obtained and the average resistance, $\Omega_{DRL}$, seen by the current is minimized with respect to the $\Omega_{SRL}$ of the SRL scheme, just for geometry and Ohm's law considerations. In appendix \ref{appA} we refer to a model explaining this effect in the particular case of point-like irradiation at the center of the basic cell.

\noindent Since the technology transfer of the DRL industrial production is difficult, a new simplified resistive scheme for high particle rate purposes has been developed. The new design, described in detail in the next section, is based on the simple single-resistive layout with a suitable surface current evacuation scheme.

\subsection{The Silver-Grid layout}
\begin{figure}
	\begin{center}
		\includegraphics[scale=0.3]{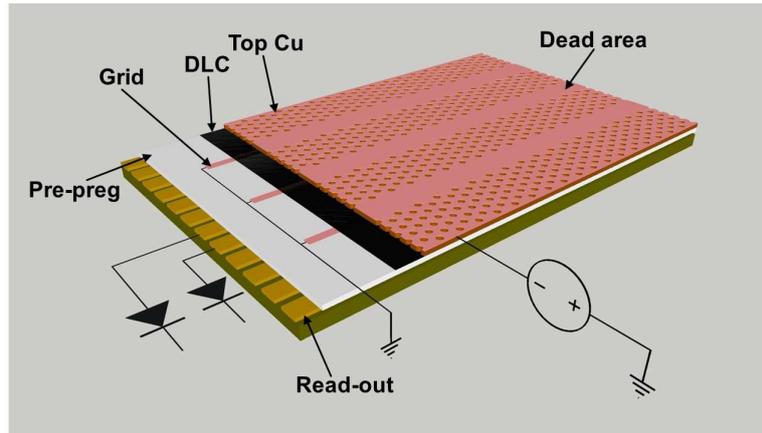}
       \caption{Sketch of the Silver-Grid layout.}
        \label{SG-sketch}
        \end{center}
\end{figure}

The Silver-Grid (SG) layout, is sketched in fig.~\ref{SG-sketch}. 
The conductive grid deposited on the DLC layer acts as a high density 2-D current evacuation system. The pitch of the grid together with the resistivity of the DLC, $\rho$, are parameters of this layout.
In the first prototypes the conductive grid has been screen-printed using a silver paste, from which the name of the layout.

\begin{figure}
	\begin{center}
		\includegraphics[scale=0.65]{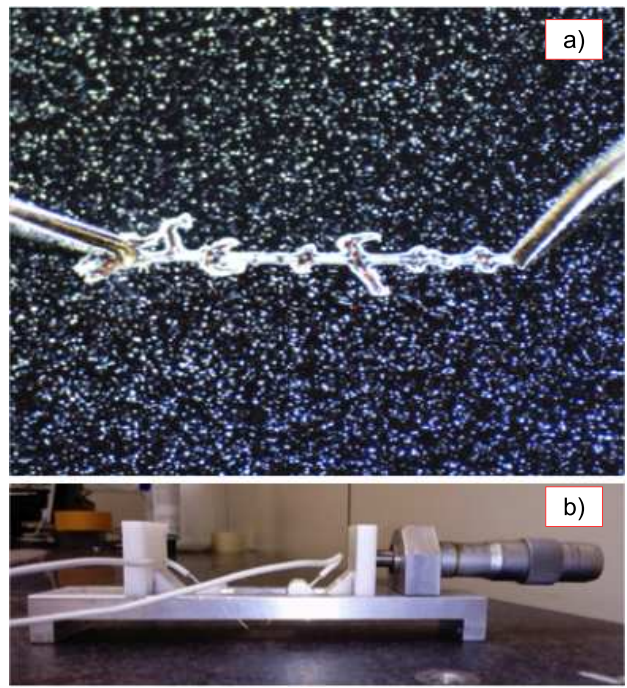}
       \caption{a)  the effect of a discharge on the DLC surface; b) the tool used for the study of the discharges on the DLC surface.}
        \label{DOCA-tool}
        \end{center}
\end{figure}

\begin{figure}
	\begin{center}
		\includegraphics[scale=0.5]{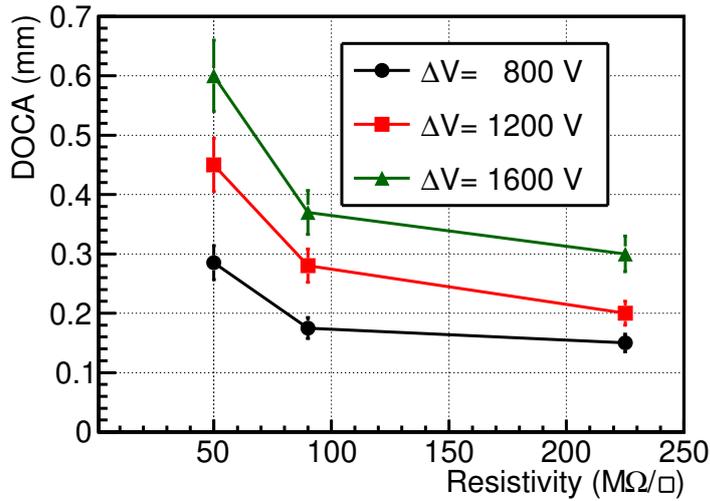}
       \caption{DOCA vs DLC resistivity for different voltages applied to the tips.}
        \label{DOCA-vs-rho}
        \end{center}
\end{figure}

\noindent The presence of a conductive grid on the DLC can induce discharges over its surface, fig.~\ref{DOCA-tool}$-a$.
This effect, depending on the resistivity, requires the introduction of a small dead zone in the amplification stage above the grid lines (see fig.~\ref{SG-sketch}). In order to properly size the dead zone, a study of the surface discharges on the DLC has been performed. For this purpose a custom tool, fig.~\ref{DOCA-tool}$-b$, composed of two movable thin conductive tips has been built. The tips are placed in contact with the DLC and supplied with HV. The distance between the two tips can be gauged with a micrometric screw.
In fig.~\ref{DOCA-vs-rho} the {\it distance-of-closest-approach (DOCA)}  before the occurrence of a discharge is reported  as a function of the DLC resistivity for different voltages.
In all the $\upmu$-RWELL layouts the DOCA can be defined as the minimum distance between a grounding line and the closest well.

\noindent In a simplified model of the resistive layouts the charge moving on the DLC would experience an effective resistance $\Omega_{eff}$ that we define as:

\begin{equation}
\Omega_{eff} = \frac{\rho}{2}\frac{(pitch/2 + DOCA)}{w}
\label{omegaeff}
\end{equation}

\begin{figure}
    \centering
    \includegraphics[scale=0.3]{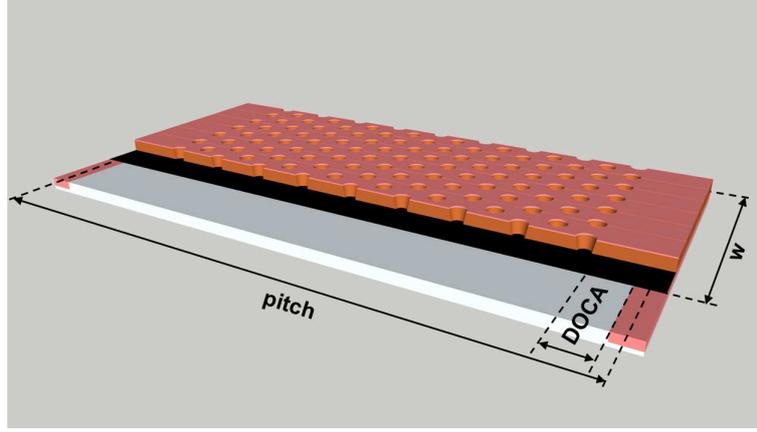}
    \caption{Cross-section of a Silver Grid layout with the definition of its parameters.}
    \label{parametri}
\end{figure}

\noindent The $\Omega_{eff}$ comes from the definition of the surface resistance, $\rho$, that is applicable in general to resistive films such as our DLC.
Considering a thin portion of the DLC (fig. \ref{parametri}), with length L and width w, the resistance is R = $\rho$ (L/w). In our estimate we have considered that w = 1 mm.
Under these assumptions, considering the approximation that the basic cell of the detector is uniformly irradiated  we obtain the $\Omega_{eff}$\footnote{ $\Omega_{eff}=\frac{\rho}{w} \left(\int_{DOCA}^{pitch/2}x dx\right) \left(\int_{DOCA}^{pitch/2}dx\right)^{-1}$}.
For the DRL layout the DOCA is the distance between a vias v1 and the closest vias v2, while the pitch is the distance between two closest vias v1 (sec. \ref{sec_DRL}).
For the SRL, being essentially a large SG, the pitch is half of the detector edge. 

\noindent It is worth pointing out how the $\Omega_{eff}$ is different from the $\Omega$ mentioned in appendix \ref{appA}: the second one actually is defined for point-like irradiation at the centre of the detector basic cell.

\noindent In table \ref{det-caracteristiche} we report the characteristics of the  HR-layouts described in this work. For completeness also the parameters of the SRL have been included.

\begin{table}
\begin{center}
\begin{tabular}{|l|c|c|c|c|c|c|} 
\hline
\textbf{Layout} & $\rho$ &  ground-pitch & dead-zone & DOCA & geometric &  $\Omega_{eff}$ \\
\textbf{} & (M$\Omega/\Box$) & (mm) &  (mm) & (mm) & acceptance (\%) & (M$\Omega$)\\
\hline
\textbf{SG1} &  70 & 6 & 2 & 0.85 & 66 & 134 \\
\hline
\textbf{SG2} &  65 & 12 & 1.2 & 0.45 & 90 & 209 \\
\hline
\textbf{SG2++} &  64 & 12 & 0.6 & 0.25 & 95 & 200\\
\hline
\textbf{DRL} &  54 & 6 & 0 & 7 & 100 & 270\\
\hline
\textbf{SRL} &  70 & 100 & 0 & 5.5 & 100 & 1947\\
\hline
\end{tabular}
\caption{Resistive and geometrical parameters of the HR-layouts compared with the low rate baseline option (SRL). For the SG models the ground-pitch is the grid-pitch.}
\label{det-caracteristiche}
\end{center}
\end{table}
 
\noindent The SG layouts exhibit a geometrical acceptance lower than the other detectors because of the presence of dead zone. Thanks to the DLC+Cu technology, allowing the photo-etching of very thin grid lines ($\simeq$100 $\upmu$m width), for the SG2++ it has been possible to reduce the dead zone down to 5\% of the active area.

\section{Performance of the HR-layouts}
The performance of the HR-layouts have been measured with pion and muon beams at the H8-SpS of CERN and the $\uppi$M1 of PSI.
The experimental set-up used in both beam tests is composed of:
\begin{itemize}
\item two couple of plastic scintillators (up-stream/down-stream), providing the DAQ trigger
\item two external triple-GEM trackers equipped with 650 $\upmu$m pitch X-Y strip read-out with analog APV25 front-end electronics \cite{apv}, defining the particle beam with a spatial accuracy of the order of 100 $\upmu$m 
\item six $\upmu$-RWELL detectors based on different resistive layouts, equipped with 0.6$\times$0.8 cm$^2$ pads and read-out with APV25 and current monitored
\end{itemize}
The gaseous detectors have been operated with Ar/CO$_2$/CF$_4$ (45/15/40) gas mixture.

\begin{figure}
	\begin{center}
				\includegraphics[scale=0.5]{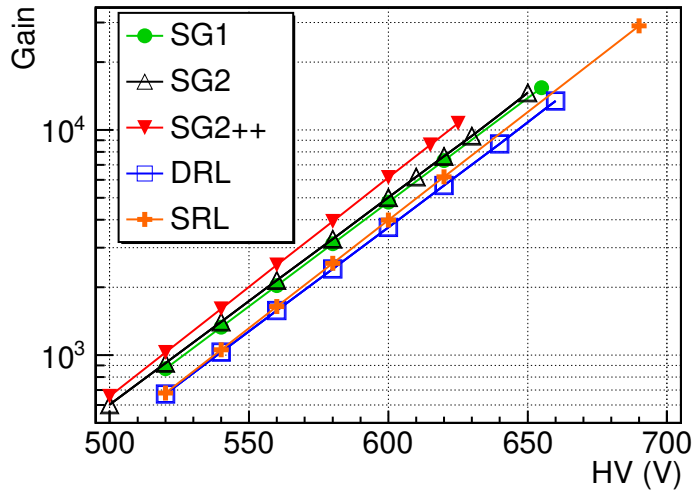}
       \caption{Gas gain of the detectors characterized at PSI. The maximum value of the gain could be correlated with the different measurement conditions in terms of particle flow.}
        \label{gain-PSI}
        \end{center}
\end{figure}

\noindent The relevant physical quantities are reported in this paper as a function of the detector gas gain  \footnote{The gain of a gaseous detector is defined as the multiplication factor of the total ionization generated by a particle crossing the gas conversion gap. Therefore it can be estimated as the ratio between the average collected charge Q and the total ionization charge.} to take into account small manufacturing differences in the amplification stage of the $\upmu$-RWELL prototypes (i.e. well diameter and shape). In fig.~\ref{gain-PSI} we plot the gas gain of all prototypes measured with a 270 MeV/c $\pi^{+}$ beam at PSI, with particle fluxes ranging from $\sim$300 kHz/cm$^2$ up to $\sim$1.2 MHz/cm$^2$ well below the onset of the Ohmic drop of the detectors gain (see sect. 4.2.1).
 
\subsection{Efficiency studies}
\begin{figure}
	\begin{center}
		\includegraphics[scale=0.5]{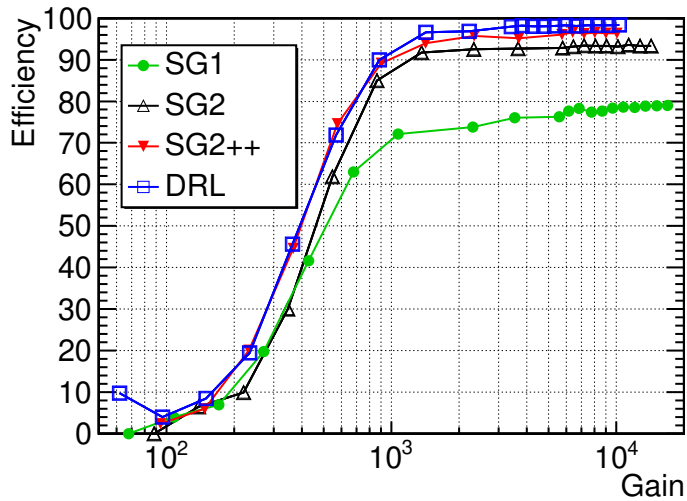}
       \caption{Efficiency as a function of the gas gain for the HR layouts.}
        \label{global-efficiency-HR}
        \end{center}
\end{figure}

\noindent In fig.~\ref{global-efficiency-HR} the efficiency of the HR-layouts is reported as a function of the detectors gain. The measurement has been performed with a flux of $\sim$300 kHz/cm$^2$ $\pi^{-}$ (350 MeV/c) and an average beam spot of 5$\times$5 cm$^2$ (FWHM$^2$). The efficiency has been evaluated considering a fiducial area of 5$\times$5 pads around the expected hit. At a gain of 5000 the DRL shows an efficiency of 98$\%$, while  the SG1 and SG2 achieve a detection efficiency of 78$\%$ and 95$\%$ respectively, larger than their geometrical acceptance. The SG2++ tends to an almost full efficiency of about 97$\%$.

\begin{figure}
	\begin{center}
		\includegraphics[scale=0.47]{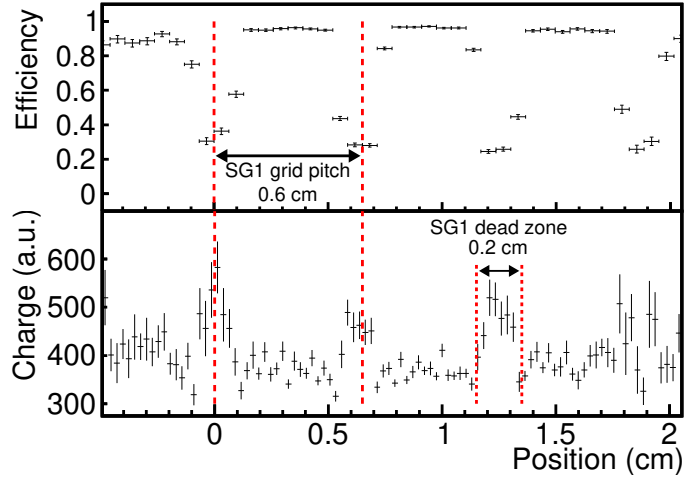}
       \caption{Efficiency and charge profile for the SG1 layout measured at CERN. Visible its geometrical features: the 2 mm wide dead zones and the 6 mm grid pitch.}
        \label{SG1-efficiency-profile}
        \end{center}
\end{figure}

\begin{figure}
	\begin{center}
		\includegraphics[scale=0.5]{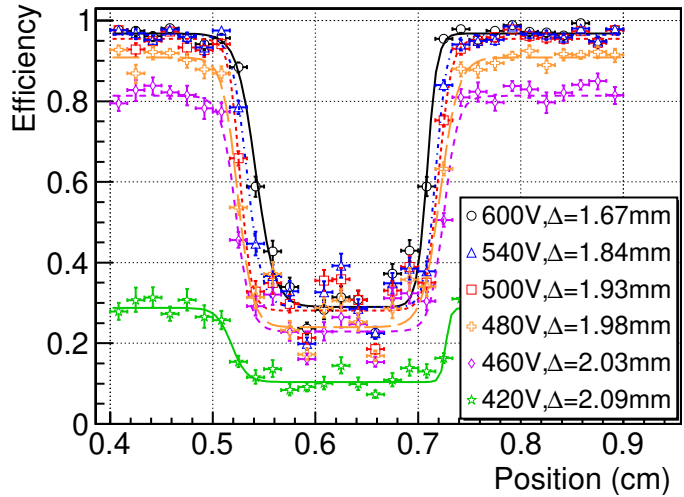}
        \caption{Zoom of the efficiency profile in the dead zone for different voltages for SG1. The $\Updelta$ is the difference between the inflection points of the two fitting Fermi-Dirac functions.}
        \label{SG2-efficiency-profile}
        \end{center}
\end{figure}

\noindent We have investigated this effect, typical of detectors with GEM-like amplification stage \cite{GEM-LHCB, RICHTER}, reporting as example the efficiency and the charge profile for the SG1. 
In fig. \ref{SG1-efficiency-profile} (top) we plot the efficiency and in fig. \ref{SG1-efficiency-profile} (bottom) the charge as a function of the incidence point reconstructed by the GEM trackers. The geometrical structure of the detector layout is visible on both: an efficiency drop and an increase of the charge are present in correspondence of the dead zones. Actually the wells close to the dead zones collect also the primary ionization produced above the inefficient regions (\emph{focusing effect}) resulting in a recovery of the detection efficiency; at the same time the amplification in these wells is increased probably due to the squeezing of the drift field lines. Moreover, as shown in fig.~\ref{SG2-efficiency-profile}, increasing the HV applied to the amplification stage the efficiency in the dead zone improves, as observed in GEM detectors (KLOE-2 CGEM \cite{kloe2}).

\subsection{Rate capability measurement}
\begin{figure}
	\begin{center}
		\includegraphics[scale=0.5]{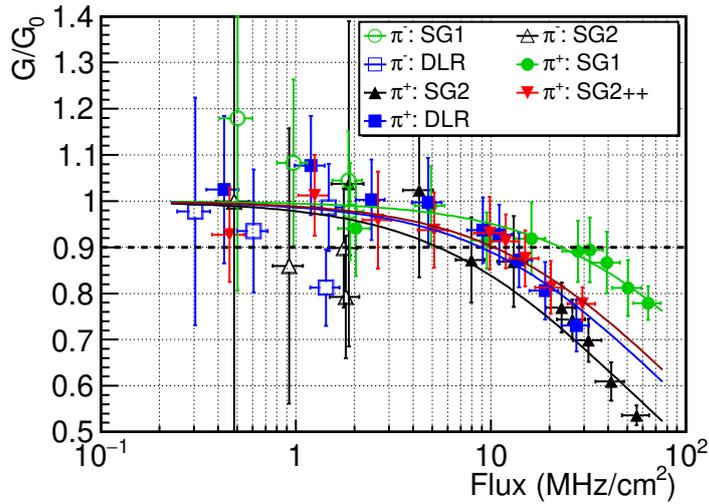}
       \caption{Normalized gas gain for the HR-layouts as a function of the pion flux. The function used to fit the points is the one derived in \cite{micro-RWELL}.}
        \label{rate-capability}
        \end{center}
\end{figure}

The rate capability of the HR-layouts has been measured at the PSI $\uppi$M1 facility that provides a quasi-continuous high-intensity secondary beam with a fluence of $\sim$10$^7$ $\pi^{-}$$/s$  and $\sim$10$^8$ $\pi^{+}$$/s$, for a momentum ranging between 270$\div$350 MeV$/c$.
This measurement can be considered reliable because the dimension of the average beam spot is larger than the basic cells of all the HR prototypes.
The result of this study is reported in fig.~\ref{rate-capability}.

\noindent The low rate measurements ($\leq$1 MHz/cm$^2$) have been performed with the $\pi^{-}$ beam, while the high intensity have been obtained with the $\pi^{+}$ beam. The detectors have been operated at a gain of about 5000.
The particle rate has been estimated with the current drawn by the GEM, that owns a linear behaviour up to several tens of MHz/cm$^2$ \cite{GEM-ELBA-2003}.
The beam spot has been evaluated with a 2-D gaussian fit of the hits reconstructed on the X-Y plane for each detector.

\noindent The gain drop observed at high particle fluxes is correlated with the ohmic behaviour of the detectors due to the DLC film. The larger the radiation rate, the higher is the current drawn through the resistive layer and, as a consequence, the larger the drop of the amplifying voltage.

\begin{figure}
	\begin{center}
		\includegraphics[scale=0.5]{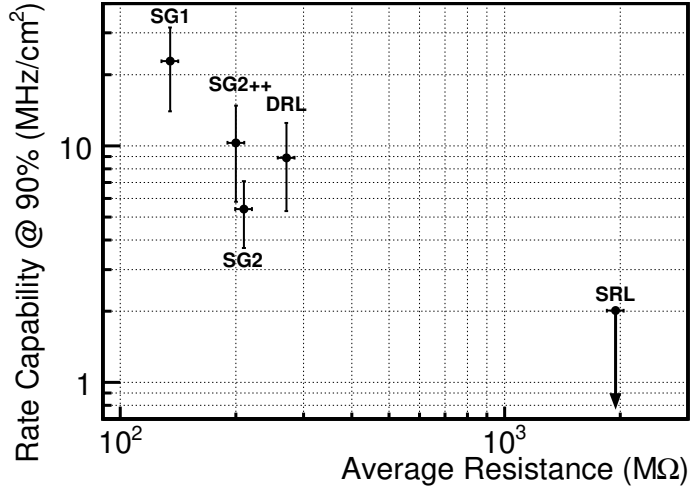}
       \caption{Rate capability at $G/G_0$ = 90\% as a function the effective resistance $\Omega_{eff}$. For the SRL, since the irradiation has not been uniform over the active area, we give an upper limit of its rate capability.}
        \label{rate-capability-2}
        \end{center}
\end{figure}

\noindent The different behaviours of the HR-layouts depend on their resistivity and current evacuation scheme and then on the $\Omega_{eff}$  (eq. \ref{omegaeff}). As shown in fig.~\ref{rate-capability-2}, the rate capability decreases as the effective resistance increases. In particular the HR-layouts corresponding to $\Omega_{eff}$ $\simeq$200 M$\Omega$ stand particle fluxes ranging from 6 to 10 MHz/cm$^2$ with high detection efficiency.

\subsection{Discharge studies}
The radiation rate at PSI allows a dedicated study on the discharges occurrence and their amplitude. We have used for the test a 270 MeV/c $\pi^{+}$ beam with a proton contamination of 3.5\%. The beam intensity has been set $\sim$90 MHz on a $\sim$5 cm$^2$ spot.

\noindent The currents drawn by all the electrodes have been recorded every second and then analyzed, removing from the analysis the values corresponding to the beam-off periods. The average current at a gain of 5000 has been of the order few $\upmu$A.

\begin{figure}[h!]
    \centering
    \includegraphics[scale=0.5]{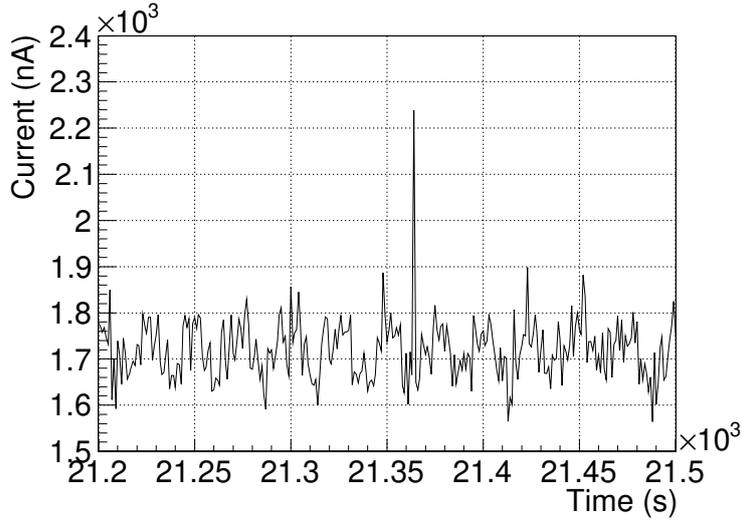}
    \caption{Typical current spike in a $\upmu$-RWELL.}
    \label{current-spike}
\end{figure}

\noindent A spark has been defined as the current spike, fig. \ref{current-spike}, exceeding 5$\upsigma$ the average current level due to the particle flux while its amplitude is the difference between the current peak and the reference current value. The spark probability per incident hadron is

\begin{displaymath}
P_{spark}=\frac{N_{spark}}{R \times \Delta t \times S}
\end{displaymath}

\noindent where $R$ is the particle rate, $\Delta t$ is the irradiation time and $S$ is the spot area
(FWHM$^2$).

\begin{figure}[h!]
    \centering
    \includegraphics[scale=0.5]{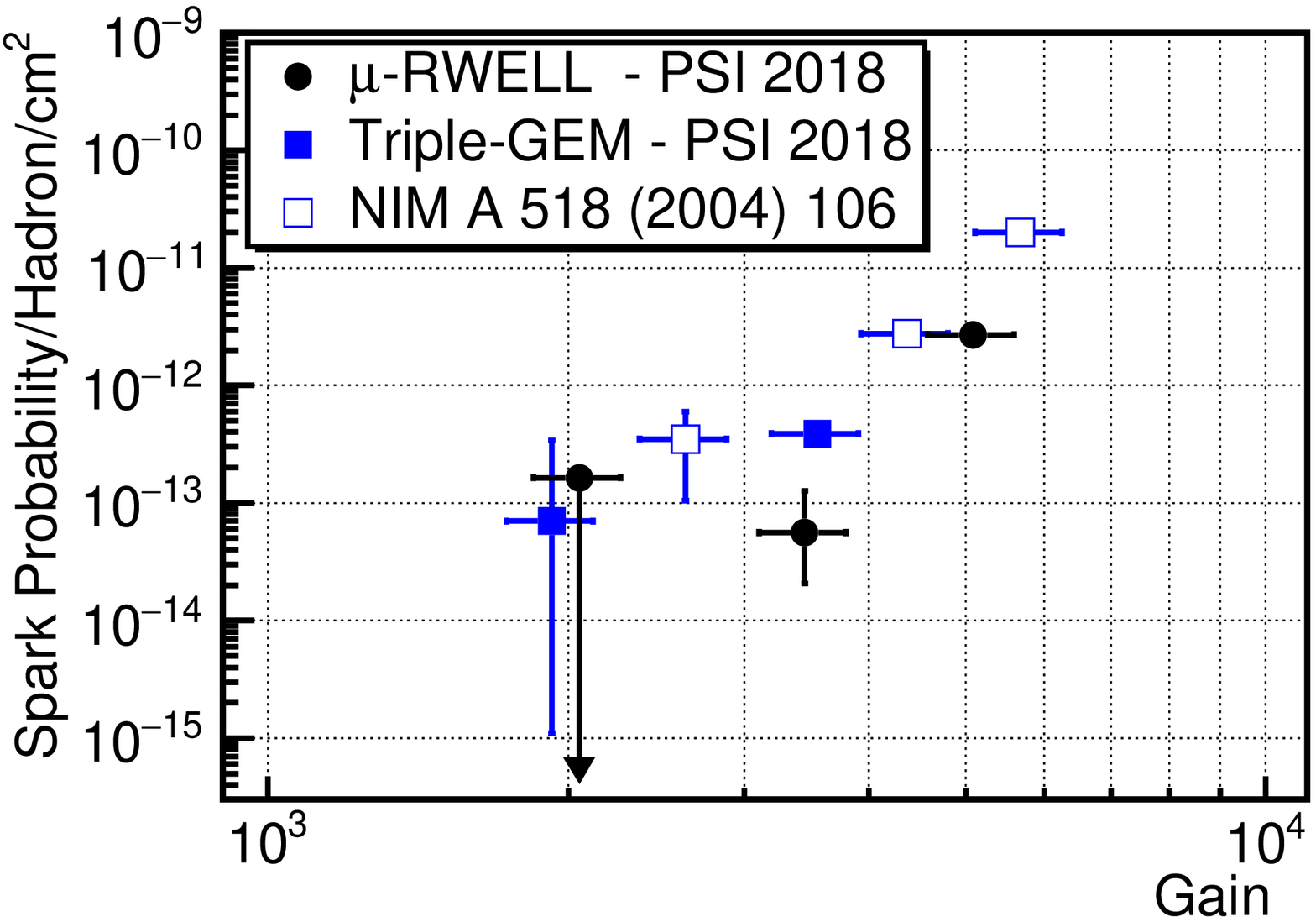}
    \caption{Spark probability per incident particle averaged on the six $\upmu$-RWELL detectors compared with GEM. The measurements at the lowest gain suffers of lack of statistics: in particular for $\upmu$-RWELL since no sparks have been detected the point represents an upper limit.}
    \label{Pspark}
\end{figure}

\noindent As shown in fig. \ref{Pspark}, the $P_{spark}$ for a $\upmu$-RWELL (full circle) is slightly lower than the one measured for triple-GEM (full square). In the plot we have reported also measurements done in the 2004 in the framework of the R\&D for LHCb (open square) \cite{GEM-ELBA-2003} with triple-GEMs flushed with the same gas mixture. These points have been renormalized to the beam conditions of 2018 by re-scaling the gain of a factor 2.29, due to the different specific ionization and to the thickness of the drift gap, and the $P_{spark}$ of a factor 0.5 because of the different proton contamination (see table \ref{beamstable}).

\begin{table}[!h]
\begin{center}
\begin{tabular}{|c|c|c|} 
\hline
 & 2004 & 2018 \\
\hline
proton contamination & 7\% & 3.5\% \\
\hline
drift gap thickness & 3 mm & 6 mm \\
\hline
beam momentum & 300 MeV/c & 270 MeV/c \\
\hline
proton specific ionization & 950 e-I/cm & 1090 e-I/cm \\
\hline
\end{tabular}
\caption{Parameters of the discharge measurements done with GEM in 2004 and in 2018.}
\label{beamstable}
\end{center}
\end{table}

\begin{figure}[!h]
    \centering
    \includegraphics[scale=0.5]{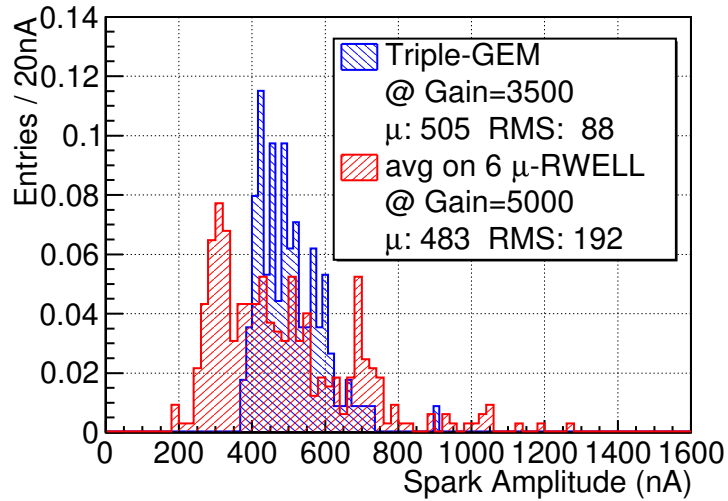}
    \caption{Spark amplitude comparison among the GEM and the six $\upmu$-RWELL detectors. The $\upmu$-RWELL have been operated at higher gain than GEM.}
    \label{dischamp}
\end{figure}

\noindent In fig. \ref{dischamp} we report the distributions of the spark amplitude measured in one of the external trackers and in the six $\upmu$-RWELLs. Taking into account the different gains (see the legend) this plot suggests that the spark amplitude for the $\upmu$-RWELL is moderately lower than for triple-GEM. 

\noindent During the irradiation test at PSI each detector has integrated a charge of the order 100 mC/cm$^2$ (in 42 hours) without showing any performance degradation. This result, confirmed by the exposure at GIF++ facility (180 mC/cm$^2$ in 250 days), should be compared with the expected 600 mC/cm$^2$ in 1 year of operation (10$^7$s) in the M2R1 HL-LHCb muon station.

\section{Conclusions}
In this paper we have discussed different resistive layouts of $\upmu$-RWELL for high rate applications. The prototypes have been tested up to particle fluxes exceeding 20 MHz/cm$^2$ at the $\uppi$M1 beam facility of the PSI.

\noindent Among the proposed layouts different solutions have been found that satisfy the very stringent requirements for the detectors in view of the phase-2 upgrades of muon apparatus at the HL-LHC as well as in the apparatus at the future accelerators FCC-ee/hh and CepC.

\noindent A rate capability up to 10 MHz/cm$^2$ with a detection efficiency of the order of 97$\div$98\% are the performance achievable with the proposed technologies.

\noindent In addition, it must be stressed that the SG2++ layout, based on DLC+Cu sputtered Apical{\textsuperscript\textregistered} foils, can be manufactured with full sequential-build-up technology, thus allowing a straightforward technology transfer of the manufacturing process to the industry.

\noindent The presence of the resistive layer helps to quench the spark amplitude, suppressing the transition from streamer to discharge.
The probability and the amplitude of the sparks for $\upmu$-RWELL result to be slightly lower than the one from triple-GEMs. 
\section*{Aknowledgements}
We would like to aknowledge Dr Zhou Yi for his contribution in preparing the DLC+Cu sputtered Apical{\textsuperscript\textregistered} foils used for the manufacturing of the improved SG2++ prototypes.
The project has been funded from the European Union's Horizon 2020 Research and Innovation program under Agreement n$^o$ 654168 - AIDA2020.

\appendix
\section{A model for HR-layouts}
\label{appA}
\begin{figure}[!h]
    \centering
    \includegraphics[scale=0.35]{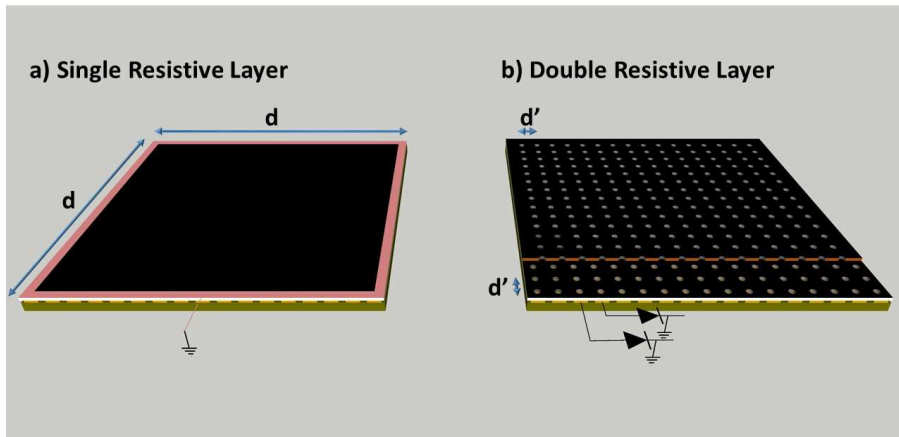}
    \caption{The charge evacuation scheme of the SRL compared to DRL.}
    \label{modello}
\end{figure}

According to the model reported in appendix B of \cite{micro-RWELL}, proposed by one of the authors (G. Morello), in a squared SRL detector of size $d$ (fig. \ref{modello}), the charge produced at its center experiences, moving toward the ground, an average resistance $\Omega_{SRL} \propto \rho d/2$. 

\noindent Assuming the same conditions, in particular the resistivity $\rho$, in a DRL detector that, whatever the size, owns a vias density of 1 cm$^{-2}$, the average resistance seen by the charge produced in the center of a matrix cell is $\Omega_{DRL} \propto \rho$ 1.5, where the factor $1.5$ averages the path on the two resistive layers.

\noindent The ratio between the two values becomes smaller as $d$ increases. In particular assuming $d'=1$ cm and $d=10$ cm, i.e. the size of the SRL used in the tests, the ratio is
\begin{displaymath}
\frac{\Omega_{DRL}}{\Omega_{SRL}}   =  \frac{\rho}{\rho} \frac{\left(d'+d'/2\right)}{d/2} = \frac{1.5}{5} = 0.3
\end{displaymath}

\noindent The rate capability depends on the inverse of the $\Omega$, so in this example for the DRL it can be expected $\sim$3 times larger than for the SRL.

\noindent To be strongly remarked that this relation stands for point-like irradiation at the center of the basic cell (i.e. X-rays) and it has been considered for the developments of the HR-layouts. 

\noindent We want to stress that this $\Omega$ is different from the $\Omega_{eff}$ cited in the text defined for a total irradiation of the basic cell.

\end{document}